\documentclass[aps, prmaterials, twocolumn, longbibliography, 10pt]{revtex4-2}

\usepackage[a4paper, total={7in, 10.3in}]{geometry}
\usepackage[utf8]{inputenc}
\usepackage[english]{babel}
\usepackage[font=small,labelfont=bf,  justification=raggedright,singlelinecheck=false]{caption} 
\usepackage{amsfonts, amsmath, amsthm, amssymb} 
\usepackage{braket}
\usepackage{graphicx}
\usepackage{footnote}
\usepackage{tabularx}
\graphicspath{{Figures/}} 
\usepackage[plainpages=false,pagebackref=false,pdftex]{hyperref}

\hypersetup{colorlinks=true,breaklinks=true,linkcolor=black,citecolor=black, urlcolor=black}
\usepackage[a-1b]{pdfx}

\begin{document}
    
\title{Graph Neural Networks for Accurate Quasiparticle Energy Levels in Large Supercells of Transition Metal Dichalcogenides}

\author{Rafael R. Del Grande}
\email{rdelgrande@ucmerced.edu}
\author{David A. Strubbe}
\affiliation{Department of Physics, University of California, Merced, California, USA}

\date{\today}

\begin{abstract}

The GW method is the one of the main workhorses of the materials science community to accurately calculate the electronic bandgap of materials, as it overcomes the gap underestimation problem from Density Functional Theory (DFT). Despite its accuracy, its computational cost becomes prohibitive when calculations have more than 10-100 atoms, which makes more difficult its use for defects, surfaces and moiré systems that demand large supercells. To overcome this problem, we developed a Graph Neural Network trained with smaller GW calculations which can make accurate predictions on large supercells at a small fraction of the computational cost.

Our model uses as features the projection of DFT wavefunctions into localized atomic orbitals, while graph nodes represent the atomic positions and edges represent atom to atom distances, which captures the system geometry. We explore how to do “small AI”, providing a workflow of how to create a personal and convenient dataset of GW calculations to train the model that can be applied to different materials so that the final user does not need to rely on massive datasets. We define what must be included in the training dataset to capture the physics of our objective simulation, and perform a Bayesian optimization of hyperparameters. In the case of monolayer MoS$_2$ with a sulfur vacancy our GNN models achieve high accuracy (MAE ranging 10 to 50 meV) with a simplified workflow applicable to other materials, accelerating accurate large electronic structure calculations.

\end{abstract}

\maketitle

\section{Introduction}

The GW approximation \cite{Hybertsen1986, Deslippe2012} is the standard method to compute quasiparticle (QP) energies from first principles. Even with recent developments in HPC \cite{Zhang2025exascale} and in the convergence of GW calculations \cite{AltmanPRL2024, JornadaPRB2017}, the computational cost of converged GW calculations is still higher than DFT computations, specially in the case of studies with defects where simulation cells need to be large enough to avoid the interaction of the defect with its own periodic images or in case of twisted multilayer materials where the cell size is given by $L=a/(2 \sin (\theta/2))$, with $a$ being the lattice parameter of one monolayer. 

Machine learning offers a way around this cost. Models that predict GW corrections from DFT inputs have been trained on large databases of bulk and two-dimensional materials with descriptors of the individual electronic states \cite{Knosgaard2022}, or with the DFT partial density of states \cite{Takano2026}; with gradient boosting on DFT eigenvalues and exchange-correlation potentials for snapshots of molecular dynamics \cite{Abdelghany2026}; by predicting the dielectric matrix \cite{Zauchner2023}; and, most recently, by predicting the full $k$-resolved self-energy matrix with a transformer \cite{Xu2026}. In parallel, equivariant graph neural networks (GNNs) predict the DFT Hamiltonian of large structures from small training cells \cite{LiDeepH2022,GongDeepHE32023}, and DeepH-GW extends this to a GW quasiparticle Hamiltonian, with transferability from small training structures to larger supercells \cite{Gong2026}. These approaches target broad chemical coverage or the full Hamiltonian, and rely on large training sets or on specialized architectures. Pant et al.\ obtained compact expressions of the QP correction in terms of DFT energies \cite{PantPRB2026}, which inspire some of the features that we test.

Here we follow a different route, which we call ``small AI'': a training set of GW calculations of the material of interest, small enough to be generated by any user, and a GNN that predicts the QP correction of each Kohn-Sham state and acts on cells of any size, pushing the limit of QP correction predictions to be the limit of DFT computations. We study monolayer MoS$_2$ with a sulfur vacancy and with a substitutional oxygen. We ask which input features carry the information, how the error depends on the training data, whether the model transfers to cells larger than any seen in training, and whether it can be used for quantities beyond the QP bands, such as electron-phonon coefficients at the GW level.


\section{Methods}

\subsection{DFT and GW}

DFT calculations were performed with the Quantum ESPRESSO code \cite{Giannozzi2009, Giannozzi2017, Giannozzi2020} and GW calculations were performed with the BerkeleyGW code \cite{Deslippe2012, Rohlfing2000, Hybertsen1986}. We used ONCV scalar-relativistic LDA pseudopotentials with standard precision from pseudodojo \cite{VansettenCompPhysCom2018, HamannPRB2013}. The unitary cell vector perpendicular to the sheet planes was chosen to be 25 \rm{\AA}. We performed G$_0$W$_0$ calculations within the Generalized Plasmon Pole \cite{Deslippe2012, Hybertsen1986}, with a cutoff energy to build the dielectric matrix equal to 25 Ry. We generated empty states with energy up to the DFT cutoff (35 Ry) using the parabands code and used the pseudo stochastic bands method to compress empty states \cite{AltmanPRL2024} with an accumulation window of 2\%, 2 stochastic pseudobands per energy subspace. We use Coulomb truncation for 2D systems in both DFT \cite{SohierPRB2017} and GW \cite{IBPRB2006} calculations. 

DFT atomic positions relaxations were performed with a cutoff energy of 80 Ry and electronic structure calculations and wavefunction generation  with a cutoff energy of 35 Ry, as the DFT bandgap converges faster than the total DFT energy \cite{AltmanPRL2024}. This choice of smaller cutoff energy for wavefunction to be used in GW calculations, saves substantial computational resources as quasiparticle energy corrections depend on double sums on $\textbf{G}$ vectors \cite{Deslippe2012}. 

For the primitive cell (PC) calculations with one Mo and two S atoms and lattice parameter equal to 3.12033 \text{\AA} we use a 6$\times$6$\times$1 k grid centered at $\Gamma$. To improve the convergence with respect to $k$-grid in GW, we used non-uniform sampling method of \cite{JornadaPRB2017}. 

\subsection{Data generation}

To generate our data, we perform calculations for primitive cells and supercells composed by 2$\times$2$\times$1, 3$\times$3$\times$1, and 4$\times$4$\times$1 replicas of PC and with kgrids 3$\times$3$\times$1, 2$\times$2$\times$1, and 2$\times$2$\times$1, respectively (see Table \ref{tab:cost_GW_SC}). For the sets with 3$\times$3$\times$1 supercells we introduce two kind of defects: sulfur vacancies and substitutional oxygen that replaces one sulfur atom (see figure xxxx). To avoid strong interactions of the defect with its periodic images in the sets with four and one primitive cells no defect is introduced. In each set we apply atomic displacements from gaussian distributions with standard deviation equal to 0.1 \text{\AA}. 

We see that even with advanced techniques to speed up GW calculations \cite{AltmanPRL2024, JornadaPRB2017}, for a large 6$\times$6 supercell their computational cost is $\sim$ 195 times the computational cost of DFT calculations (see Table \ref{tab:cost_GW_SC}, which makes many-body studies for those systems computationally prohibitive. In this work we propose a Graph Neural Network trained in small GW calculations that predicts quasiparticle energy levels of large systems.

\begin{table*}[t]
\caption{GW data sets and GPU node-hours used in Perlmutter-NERSC of the production GW workflow per replica. Number of atoms ($N_a$), number of valence electrons $N_e$ (Mo 14 , S and O 6), $k$-grid of the DFT and GW calculations, number of replicas with finished GW and number of $\Sigma$ bands per $k$-point, followed by the cost of each stage. The $\Sigma$ stage is given per state (band and $k$-point; 252 states for $1\times1$, 2$\times$2 and for $3\times3$, 160 for $4\times4$) and in total.  $^\dagger$: For $4\times4$ + S vac only, some calculations ran in other cluster (Frontera-TACC) so we assume its computational costs to be the same as 4$\times$4 O$_{\rm{S}}$. The $6\times6$ estimate (108 atoms, no defect, $\Gamma$ only, 40 $\Sigma$ states) evaluates log-log power-law fits of each step against $N_e$ through the $1\times1$, $2\times2$, $3\times3$ + O$_{\rm S}$ and $4\times4$ + O$_{\rm S}$ cells (exponents 1.16 for DFT, 1.51 for parabands, 1.56 for $\epsilon$, 2.35 for $\epsilon$ NNS and 1.33 for $\Sigma$ per state). A GNN-based workflow needs only the DFT stage.}
\label{tab:cost_GW_SC}
\footnotesize
\setlength{\tabcolsep}{4pt}
\begin{tabular}{lrrcccrrrrrrr}
\hline\hline
 & & & & & & \multicolumn{7}{c}{GPU node-hours} \\
\cline{7-13}
Cell & $N_a$ & $N_e$ & $k$-grid & Replicas & $\Sigma$ bands & DFT & parabands & epsilon& epsilon NNS& $\Sigma$ per state & $\Sigma$ total & Total \\
\hline
    $S_1=$$1\times1$& 3 & 26 & $6\times6$ & 25 & 7 & 0.02 & 0.04 & 0.10 & 0.03 & 0.0038 & 0.95 & 1.14 \\
    $S_2=$$2\times2$& 12 & 104 & $3\times3$ & 25 & 28 & 0.02 & 0.24 & 0.62 & 0.69 & 0.0182 & 4.59 & 6.17 \\
    $S_{3,v}=$$3\times3$ + S vac & 26 & 228 & $2\times2$ & 5 +1 relaxed & 63 & 0.04 & 0.79 & 1.73 & 4.28 & 0.0442 & 11.13 & 17.97 \\
    $S_{3,o}=$$3\times3$ + O$_S$ & 27 & 234 & $2\times2$ & 5 + 1 relaxed & 63 & 0.26 & 0.78 & 1.71 & 4.30 & 0.0426 & 10.74 & 17.80 \\
    $S_{4,v}$$=$$4\times4$ + S vac & 47 & 410 & $2\times2$ & 1 relaxed & 40 & 0.47$^\dagger$ & 3.33$^\dagger$ & 8.91$^\dagger$ & 22.45$^\dagger$ & 0.1936 & 30.98 & 66.14$^\dagger$ \\
    $S_{4,o}$$=$$4\times4$ + O$_S$ & 48 & 416 & $2\times2$ & 1 relaxed & 40 & 0.47 & 3.33 & 8.91 & 22.45 & 0.1839 & 29.43 & 64.59 \\
\hline
    $6\times6$ (estimate) & 108 & 936 & $\Gamma$ & -- & 90& $\sim$ 0.9& $\sim$8& $\sim$21.8& $\sim$130& $\sim$0.38& $\sim$ 34 & $\sim$195\\
\hline\hline
\end{tabular}
\end{table*}

\section{Correlations in GW energy levels}

\begin{figure}
    \centering
    \includegraphics[width=\linewidth]{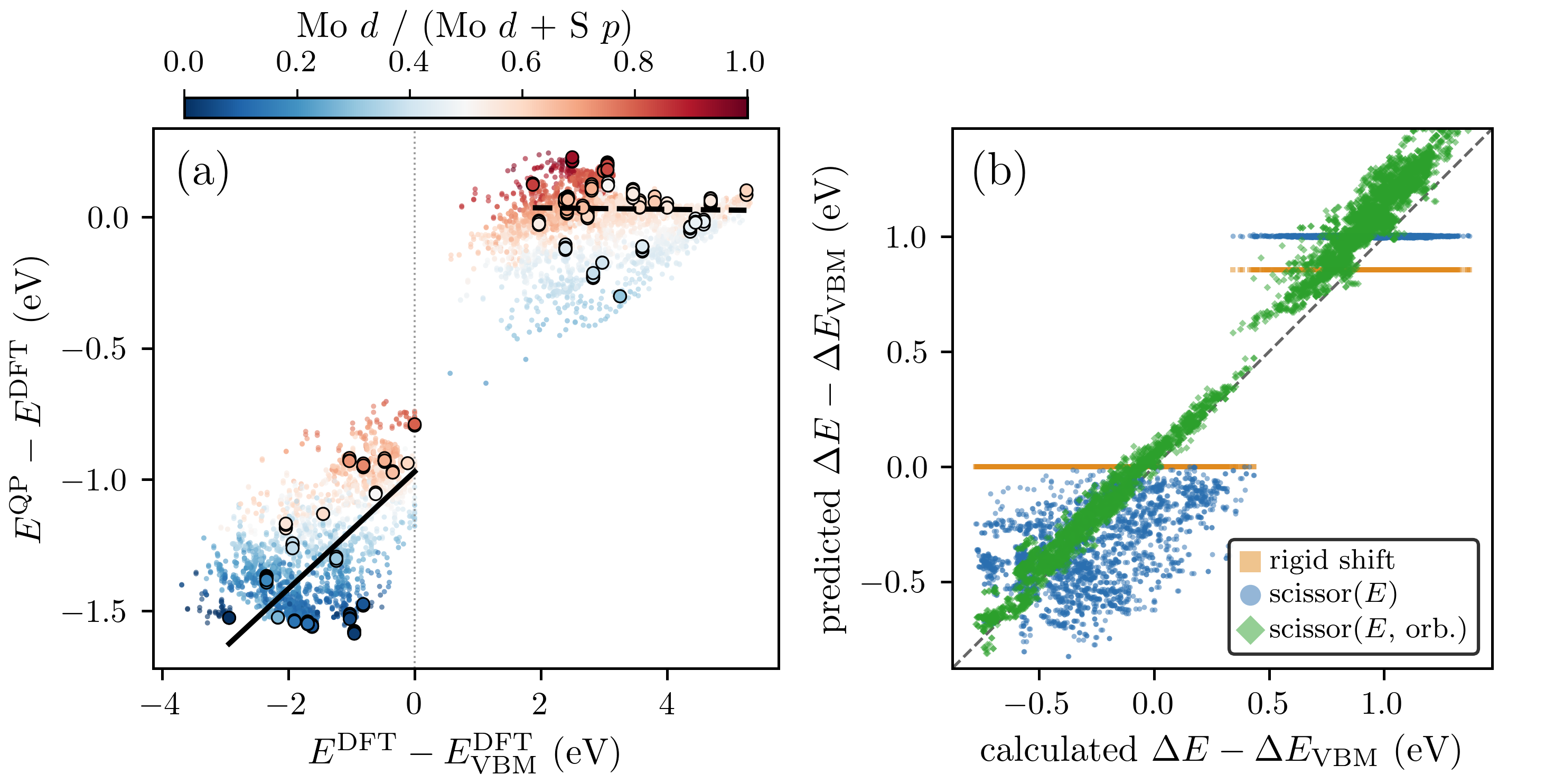}
    \caption{Quasiparticle correction $\Delta E=E^{\rm QP}-E^{\rm DFT}$ of the $1\times1$ MoS$_2$ cell.
(a) $\Delta E$ versus $E^{\rm DFT}-E^{\rm DFT}_{\rm VBM}$ for the equilibrium structure (large circles with black edge; 7 bands) and for the 25 displaced structures (small circles); color: Mo-$d$ fraction $x=$ Mo\,$d$/(Mo\,$d$+S\,$p$) of the state. The lines are the scissor operator linear in $E^{\rm DFT}$, fitted only to the equilibrium structure separately for the valence (solid, $\Delta E=0.22\,E-0.97$ eV) and conduction states (dashed, $\Delta E=-0.00\,E+0.04$ eV). Two further models fitted on the equilibrium data are used: the scissor operator that also depends on the orbital character, $\Delta E=\alpha E+\beta x+c$ (valence $\alpha=0.03$, $\beta=0.89$ eV, $c=-1.54$ eV; conduction $\alpha=0.04$, $\beta=0.77$ eV, $c=-0.57$ eV), and a rigid shift of the DFT conduction states by $0.86$ eV, which makes the DFT minimum gap equal to the GW one. (b) $\Delta E$ predicted by the three models of (a) against the calculated one for the displaced structures, both referred to the correction at the VBM of each structure.
Mean absolute errors for valence/conduction states: rigid shift 310/148 meV, scissor($E^{\rm DFT}$) 195/137 meV, scissor($E^{\rm DFT}$, orbital) 61/142 meV.}
    \label{fig:scissor_operator_1PC}
\end{figure}

\begin{figure}[h!]
    \centering
    \includegraphics[width=\linewidth]{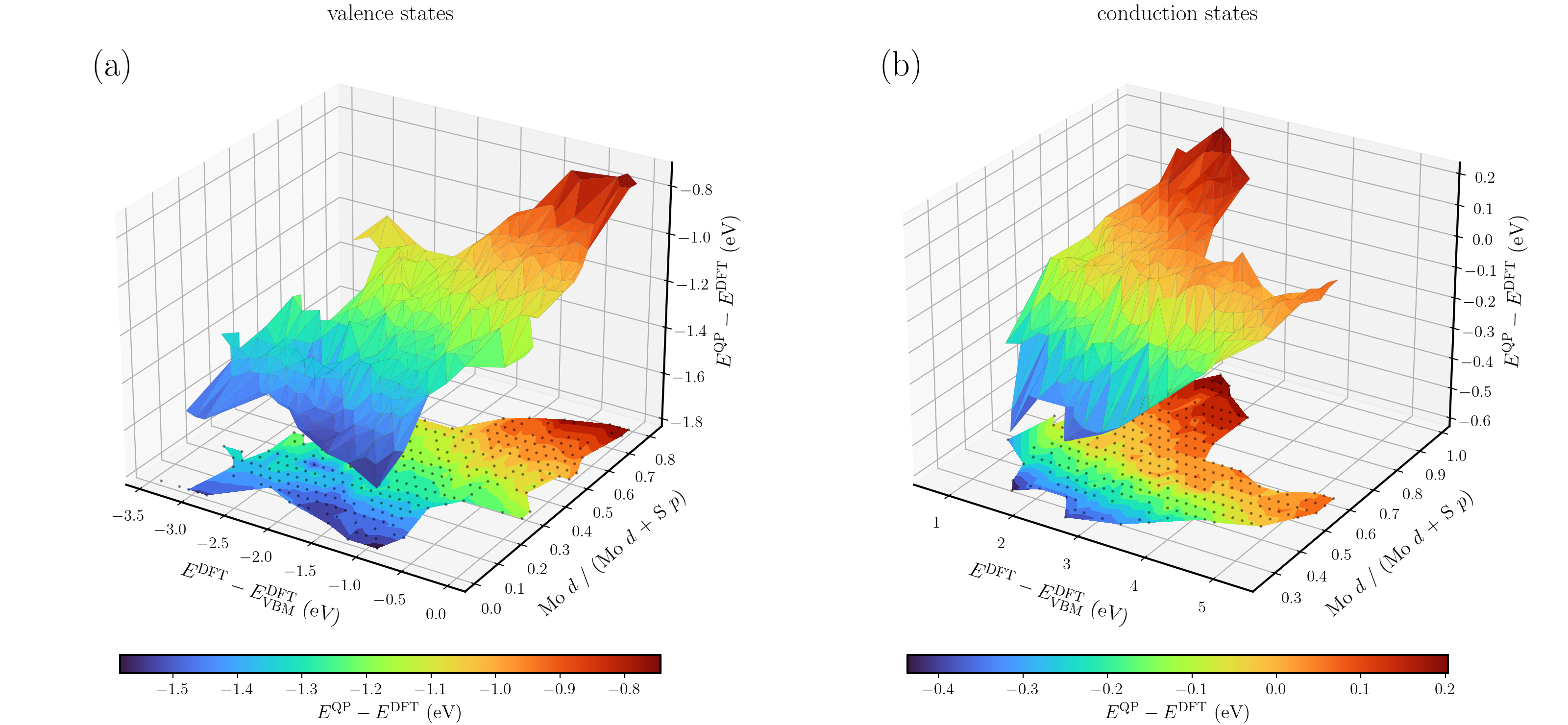}
    \caption{Quasiparticle correction $\Delta E=E^{\rm QP}-E^{\rm DFT}$ of the $1\times1$ MoS$_2$ cell as a function of the DFT energy and of the orbital character of the state, for (a) the three highest valence bands and (b) the four lowest conduction bands in the GW window. The data comes from the 25 displaced structures in the primitive cell.}
    \label{fig:3d_DeltaE_vs_Edft_orb_1PC}
\end{figure}

The quasiparticle (QP) energy levels in the GW approximation are given by 

\begin{equation}
    E^{\rm{QP}}_{n\textbf{k}} = E^{\rm{DFT}}_{n\textbf{k}} + \langle n \textbf{k} | \Sigma(E^{\rm{DFT}}_{n\textbf{k}}) - V_{xc}|n \textbf{k} \rangle
\end{equation}
where in the so called G$_0$W$_0$ calculations we approximate the GW and DFT Hamiltonians share the same eigenvectors $|n\textbf{k} \rangle$. Two very common approximations are the rigid shift, where one sums an energy $\Delta$ to the DFT energy levels to make the bandgap to match the QP or some experimental value for the gap, and scissor operators where one assumes that the valence (conduction) QP corrections depend linearly on the DFT energies. In figure \ref{fig:scissor_operator_1PC} we compare the performance of rigid shift, scissor operators dependent on DFT energies ($\Delta E = \alpha E^{\rm{DFT}}+c$) and scissor operators dependent on DFT energies and on projection $x= |\langle \psi_{n\textbf{k}}|d_{\rm{Mo}}\rangle|^2 / (|\langle \psi_{n\textbf{k}}|d_{\rm{Mo}}\rangle |^2 + \langle |\psi_{n\textbf{k}}|p_{\rm{S}}\rangle |^2)$, where x = 1 (0) represents that the state is totally composed by the orbital $d$ ($s$) of atom Mo (S). We fit those three models using the relaxed atomic positions PC data and apply them to predict the PC data with randomly displaced atoms. The rigid shifts have a bad performance in general, while the energy dependent scissor operator has a bad performance in general for the conduction states. The inclusion of projection in the fitted model decreases the mean absolute error (MAE) from 195 to 61 meV for valence states and for conduction states the error goes from 137 to 142 meV. In Figure \ref{fig:3d_DeltaE_vs_Edft_orb_1PC} we show the dependence of the QP energy corrections on the DFT energies and orbital character. From those results we can see that the relation between those quantities is more complex than linear dependence.

\section{Graph Neural Networks}

\label{sec:graph_neural_networks}

A graph neural network (GNN) acts on a graph whose nodes carry feature vectors and whose edges carry geometric information. In each layer every node collects messages from its neighbors, with weights that depend on the edge, and updates its own features; stacking layers lets information propagate over several bonds, and a final pooling step produces a graph-level prediction \cite{gilmer2017neural,battaglia2018relational}. The construction is invariant under permutations of the atoms, translations and rotations of the structure when the edge features are built from distances and angles. GNNs have become standard for molecules and crystals \cite{schutt2018schnet,xie2018crystal}, with distances entering through radial filters \cite{simonovsky2017dynamic} and bond angles through triplet messages \cite{klicpera2020directional}. 

First we build the graph that represents the atomic positions $R$, with its nodes representing atomic positions and edges representing the connection between two atoms. We expand each DFT state in atomic orbital basis $|\psi_{n\textbf{k}}(R)\rangle = \sum_{o,R_A} p_{R_A,o}^{n \textbf{k}} | \phi_{o}(r-R_A) \rangle$ , where $R_A$ is the position of atom $A$ and $o$ is the atomic orbital (1s, 2s, 2p, ...) and $p_{R_A,o}^{n\textbf{k}}=\langle \phi_{o}(r-R_A)|\psi_{n\textbf{k}} \rangle$. For our most basic model, each DFT state will have a set of scalars to characterize it: its occupation, being 0 for unoccupied (conduction) or 1 for occupied (valence) state and its DFT energy in relation to the Fermi energy. Nodes are the atoms and edges connect atoms within a cutoff radius 4 \text{\AA} from the hyperparameter search (Sec. \ref{sec:hp}) with periodic boundary conditions, and every pair of bonds sharing an atom forms an angle. A layer is an edge-conditioned convolution \cite{simonovsky2017dynamic}: a small multilayer perceptron (MLP) maps the Gaussian-expanded bond length (16 radial functions, damped by a smooth cosine envelope at the cutoff) and any further edge features to the weight matrix applied to the neighbor's features. An angle branch adds a contribution from every triplet, gated by the features of its two end atoms. Each layer is followed by batch normalization, a ReLU activation and dropout. After three layers the atom features are pooled with weights proportional to the total projected weight of the state on each atom (so atoms where the state is concentrated dominate), and a readout MLP maps the pooled vector, together with state-level scalars, to $\Delta E$. The target is the raw QP corrections.

We also have tested other kind of features that can be included in our GNN model

\begin{itemize}
    \item Inverse participation ratio (IPR): the IPR is defined as $\sum_i^N |\langle \psi_{n\textbf{k}}|\phi_i \rangle |^4$ for a normalized state $\psi_{n\textbf{k}}$. For a complete localized state it is 1, while for a totally delocalized it is $1/N$ 
    \item DFT energies: for each state we can compute the kinetic, Hartree, exchange-correlation and external potential energies from DFT wavefunctions and the electron density. Pant et al. use those energies to obtain simple expressions of the QP energy corrections as function of those DFT energies \cite{PantPRB2026}.
    \item  Orbital edge features: bond projected quantities built from the complex orbital amplitudes of the state, namely the $\sigma$ and $\pi$ weights of the $p$ shell and the $\sigma$, $\pi$ and $\delta$ weights of the $d$ shell along each bond. They describe how the orbitals of the state are oriented with respect to the bonds.
\end{itemize}

\section{Hyperparameters choice}

The models were built with PyTorch \cite{paszke2019pytorch} and PyTorch Geometric \cite{fey2019fast}. The architecture and optimizer hyperparameters were chosen by Bayesian optimization \cite{snoek2012practical,shahriari2016taking}, as implemented in Optuna \cite{akiba2019optuna} (Optuna 4.9.0). Table \ref{tab:hp} lists the search space. Each trial builds the graphs at the sampled edge cutoff, trains for a fixed 75 epochs with the Adam optimizer \cite{kingma2015adam} and the mean squared error loss (batch size 32, no pruning and no early stopping inside a trial), and returns the mean absolute error (MAE) on the \emph{dev} structures after the last epoch. Each search was run for 80 trials.

\begin{table*}[t]
\caption{Hyperparameter search space and selected values (HP S3).}
\label{tab:hp}
\begin{tabular}{lll}
\hline
Hyperparameter & Search space & Selected \\
\hline
Edge cutoff (\AA) & \{3.0, 3.5, 4.0\} & 4.0 \\
Learning rate (Adam) & $10^{-3}$--$10^{-1}$, log-uniform & $4.2\times10^{-3}$ \\
Message-passing layers & 2, 3 & 3 \\
Width of each layer & 32--256, step 16 (independent per layer) & 176, 80, 192 \\
Hidden width of the readout MLP & 64--256 & 190 \\
Dropout rate & 0--0.4 & 0.22 \\
\hline
\end{tabular}
\end{table*}

The data sets used are specified in table \ref{tab:cost_GW_SC}. The groups $S_1$, $S_2$, $S_{3,v}$ and $S_{3,o}$ are split into training, development and validation (17/4/4 for $S_1$ and $S_2$, 6/2/2 for $S_{3,v}$ and 3/1/1 for $S_{3,o}$). The dev structures drive the search, the early stopping and the learning-rate schedule; the validation structures are never used for any choice, which avoids the selection bias that arises when a model is scored on the data used to select it \cite{cawley2010overfitting}. The larger supercells ($4\times4$) serve only as tests.

With the selected hyperparameters, every model is trained with Adam and the mean squared error (MSE) loss, batch size 32, for at most 200 epochs. The learning rate is reduced by a factor 0.8 when the dev loss has not improved for 10 epochs (minimum $10^{-6}$), training stops when the dev MAE has not improved for 25 epochs \cite{prechelt1998early}, and the weights of the epoch with the lowest dev MAE are kept. Dropout \cite{srivastava2014dropout} with the selected rate is the only other regularization to avoid overfitting. Each configuration is trained with five random seeds, and results are reported as mean and standard deviation over seeds.

\section{Training process}

\begin{figure*}
    \centering
    \includegraphics[width=\linewidth]{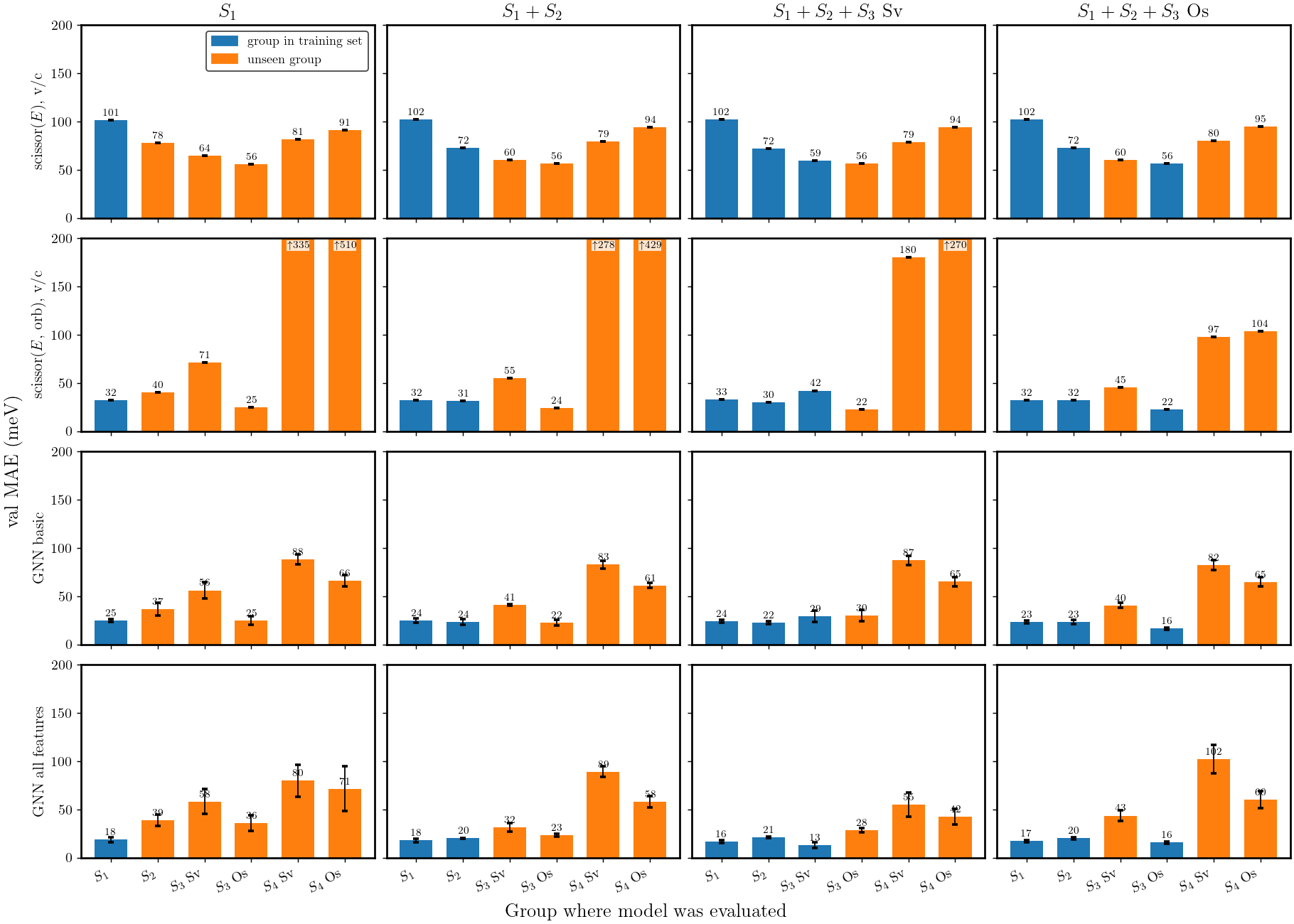}
    \caption{Scissor operators compared with the GNN. Mean absolute error on the held-out validation (val) sets of each evaluation group ($S_1$, $S_2$, $S_{3,v}$, $S_{3,o}$, $S_{4,v}$ $S_{4,o}$), for models trained on $S_1$, $S_1+S_2$, $S_1+S_2+S_{3,v}$ and $S_1+S_2+S_{3,o}$ (columns). Blue bars are groups included in the training set and orange bars are unseen groups. Rows: scissor operator $\Delta E_{c(v)}=\alpha(E^{\rm DFT}_{c(v)}-E_{\rm F})+c$; the same with an added term $\sum_{\rm orb}p_{\rm orb}\beta_{\rm orb}$, where $p_{\rm orb}$ is the projected weight on the Mo $s$, $p$, $d$ and S $s$, $p$ orbitals (summed over atoms, degeneracy-averaged); and the GNN with the basic feature set and with all features (mean over 5 seeds, error bars the standard deviation). Bars above 200 meV are cut at the top of the panel and labeled with their value. Without orbital information the scissor error is 56--102 meV on all groups; with it, 22--42 meV on groups in or near the training set, but 97--510 meV on the $4\times4$ cells.}
    \label{fig:train_eval_modes}
\end{figure*}

In Figure \ref{fig:train_eval_modes} we show the performance of scissor operators, our most basic GNN model and our GNN model including IPR, DFT energies and edge correlations (see section \ref{sec:graph_neural_networks}). We vary the training datasets and check the performance of each trained model into different groups. For the GNNs and the scissor operator with orbital projections the MAE decreases substantially if evaluated data belongs to the group in the training data set (blue bars in Figure \ref{fig:train_eval_modes}), while the generalization for larger supercells is better for GNNs trained with $S_1$, $S_2$, and $S_{3,v(o)}$ groups with MAEs down to $\sim$ 50 meV. 

\section{GW energy levels MoS$_2$ with defects}

\begin{figure}
    \centering
    \includegraphics[width=\linewidth]{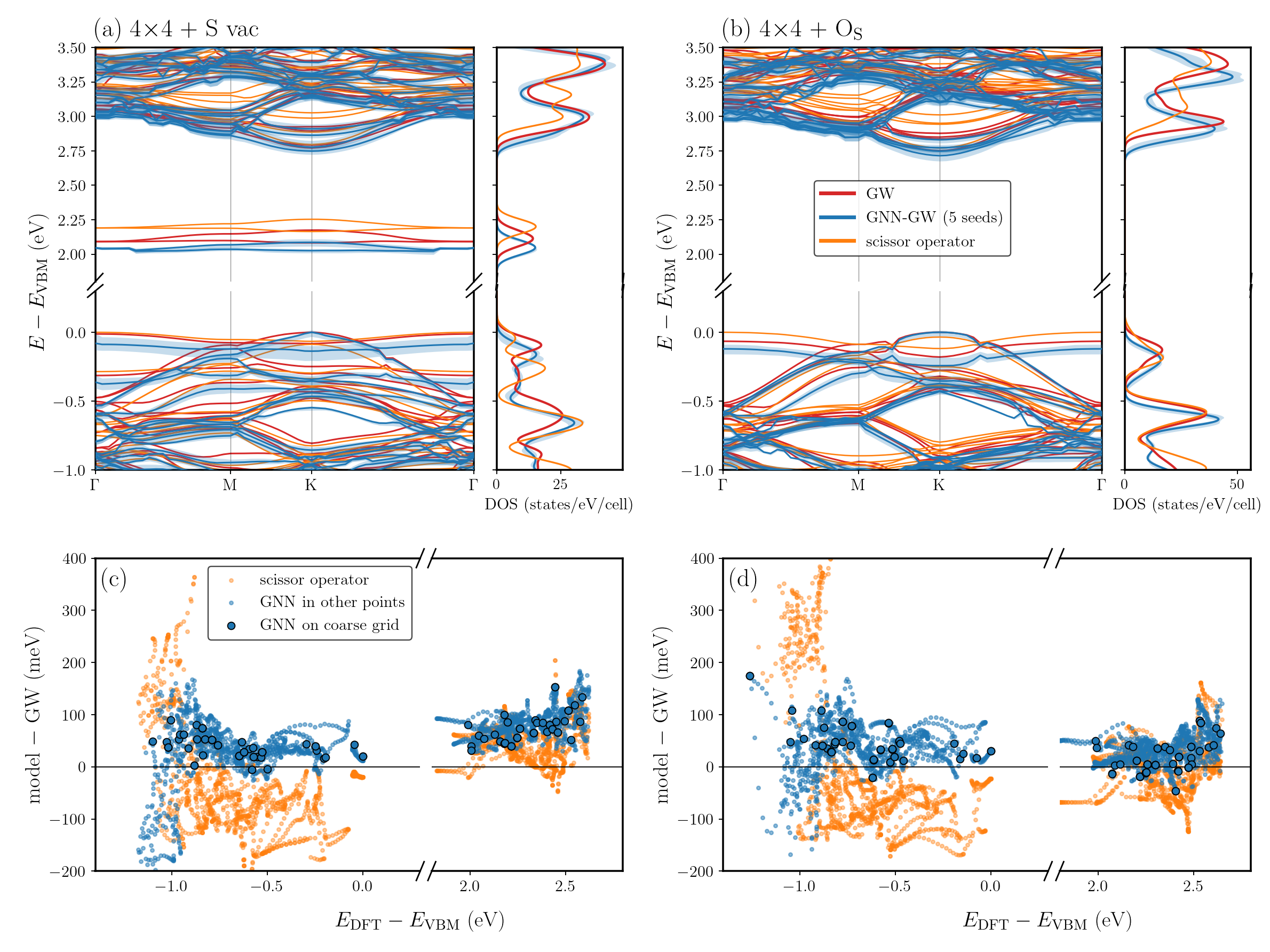}
    \caption{GW, GNN-GW and scissor-operator quasiparticle energies of the relaxed $4\times4$ MoS$_2$ cells with an S vacancy (left column) and with an O atom replacing S (right column). (a),(b) Bands along $\Gamma$--M--K--$\Gamma$ and the density of states. GW (red) values are interpolated from the $2\times2$ grid to the path. The GNN-GW energies (blue; mean of 5 seeds, shading is the range over seeds) are $E^{\rm DFT}$ plus the predicted correction; the scissor operator (orange) is $\Delta E=\alpha(E^{\rm DFT}-E_{\rm F})+c$ with separate fits for valence and conduction states. The GNN (all features) and the scissor were trained on the $1\times1$, $2\times2$ and $3\times3$ S-vacancy structures only, so neither has seen any $4\times4$ cell.
(c),(d) Difference between each model and GW for every state of the path as a function of the DFT energy relative to the DFT valence-band maximum: scissor operator (orange), GNN at the other path points (small blue dots) and GNN at the coarse-grid points $\Gamma$ and M, where the GW reference is not interpolated (blue circles with black edges). The mean absolute errors are GNN 64 meV and scissor 69 meV for the S vacancy, and GNN 42 meV and scissor 76 meV for O$_{\rm S}$; the mean errors are $+50$ and $+34$ meV for the GNN and $+7$ and $-2$ meV for the scissor.}
    \label{fig:bands_sc_4x4}
\end{figure}

Then we compute the GW bands for $4\times 4$ supercells with a sulfur vacancy and a substitutional oxygen (see Figure \ref{fig:bands_sc_4x4}). We compare the GNN with the scissor operator $\Delta E=\alpha(E^{\rm DFT}-E_{\rm F})+c$, fitted separately for valence and conduction states on the same training structures, which is the simple baseline that needs no orbital weights (Sec.~\ref{sec:gw_corr}).

In both cells the GNN reproduces the main features of the GW spectrum: the band edges, the dispersion of the valence and conduction manifolds and the density of states. For the S vacancy it places the group of empty defect states in the gap, about 2.05~eV above the valence-band maximum, within $\sim$50~meV of the GW levels, while the scissor operator, which only sees the DFT energy, puts them $\sim$100~meV too high (Fig.~\ref{fig:bands_sc_4x4}a). The mean absolute error over the states of the path is 64~meV for the GNN and 69~meV for the scissor operator for the vacancy, and 42 and 76~meV for O$_{\rm S}$. 

\section{Electron-phonon coefficients}

To confirm the efficiency of our GNN model we use it to compute electron-phonon coefficients using finite differences. Those could be computed using the GW perturbation theory (GWPT) method \cite{ZhengluPRL2019}, but its computational cost increases greatly compared to GW calculations. For the $A_1'$ mode of the $1\times1$ cell with frequency equal to 412 cm$^{-1}$, in which the two S atoms move in opposite directions perpendicular to the plane, we compute the electron-phonon coefficient $g^{\rm{QP}}_{A_1', n\textbf{k}}=\partial E^{\rm{QP}}_{n\mathbf{k}}/\partial d_{\rm{S-S}}$ using finite differences for both GW and GNN predictions. We also compare with results from density functional perturbation theory (DFPT) and DFT $g^{\rm{DFT}}_{A_1', n\textbf{k}}=\partial E^{\rm{DFT}}_{n\mathbf{k}}/\partial d_{\rm{S-S}}$. The difference between both is the matrix element $\langle n\textbf{k}|\partial (\Sigma -V_{xc})|n\textbf{k}\rangle$ which includes derivatives of the self-energy operator. For the GNN this requires no new GW calculation: the DFT eigenstates and wavefunctions of the structures displaced by $\pm \delta d_{\rm{S-S}}$ are the input of the network, and the derivative computed with finite difference of the predicted energies. 

\begin{figure}
    \centering
    \includegraphics[width=\linewidth]{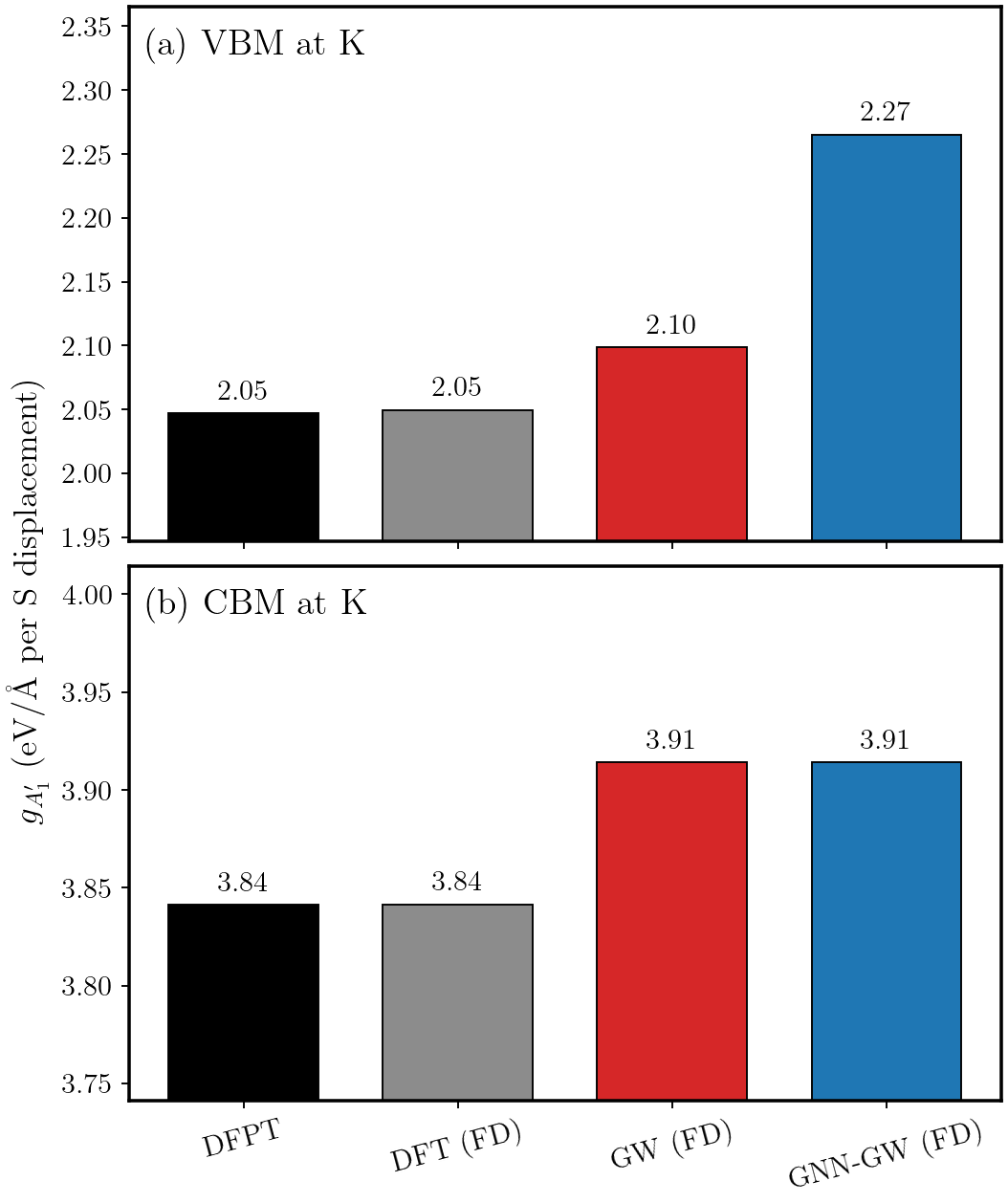}
    \caption{Electron-phonon coupling $g_{A_1'}=\partial E/\partial d$ of the $A_1'$ mode of the $1\times1$ MoS$_2$ cell, in eV/\text{AA} per S displacement (the top S atom is moved by $+d$ and the bottom one by $-d$ along the direction perpendicular to the MoS$_2$ plane), for (a) the valence-band maximum and (b) the conduction-band minimum at K. Calculations done with density functional perturbation theory (DFPT) and finite differences (FD, $d=0.04$ \AA) with DFT, GW and GNN-GW energies.}
    \label{fig:elph_GW}
\end{figure}

Figure~\ref{fig:elph_GW} shows the coupling at K. The finite differences with DFT energies reproduce DFPT (2.05 and 3.84~eV/\AA\ for the valence-band maximum and the conduction-band minimum), showing that displacements of xx were sufficiently small for the convergence of finite differences. The GW correction increases both couplings slightly, by 2\% (2.10 and 3.91~eV/\AA). The GNN-GW finite differences agree with GW for the conduction band (3.91~eV/\AA) but overestimate the valence-band coupling (2.27~eV/\AA, 8\% above GW).

\section{Conclusions}

We have trained graph neural networks on a small, purpose-built set of GW calculations of monolayer MoS$_2$ (primitive cell, $2\times2$ and $3\times3$ supercells with a sulfur vacancy and with a substitutional oxygen) to predict the GW quasiparticle correction of Kohn-Sham states. The inputs are the projections of the DFT states on atomic orbitals, the DFT energy and the occupation, and the geometry through bond lengths and angles; the hyperparameters were chosen by Bayesian optimization and the results are averages over five seeds. The main findings are: (i) for groups in the training set the error is 10--20~meV, much lower than that of the rigid shift and the scissor operators of Sec.~\ref{sec:gw_corr}; (ii) for the unseen $4\times4$ cells with a vacancy or an oxygen substitution the error is 40-65~meV and, unlike the scissor operator, the GNN reproduces the defect levels and the dispersion of the bands; (iii) the model can be differentiated by finite differences to give GW-level electron-phonon coefficients, in agreement with GW for the conduction-band coupling at K of the $A_1'$ mode and with an 8\% overestimate for the valence band. 

\section*{Acknowledgments}

R.R.D.G. and D.A.S. were supported by the U.S. National Science Foundation under Grant No. DMR-2144317. Computational resources were provided by the National Energy Research Scientific Computing Center (NERSC), a U.S. Department of Energy Office of Science User Facility operated under Contract No. DE-AC02-05CH11231; the Texas Advanced Computing Center (TACC) at The University of Texas at Austin (http://www.tacc.utexas.edu); and the Pinnacles cluster (NSF MRI, \# 2019144) at the Cyberinfrastructure and Research Technologies (CIRT) at the University of California, Merced.


\bibliography{refs}

\section{Appendix}

\end{document}